\documentclass[runningheads]{llncs}

\usepackage[pdftex]{graphicx}
\usepackage[cmex10]{amsmath}

\usepackage{algorithm}
\usepackage{algorithmicx}
\usepackage{algpseudocode}

\usepackage{array}
\usepackage{colortbl,booktabs} 

\usepackage{mdwmath}
\usepackage{mdwtab}
\usepackage{subfig}
\usepackage{eqparbox}
\usepackage{fixltx2e}
\usepackage{stfloats} 
\usepackage{url}
\usepackage{amsmath} 
\usepackage{amssymb}
\usepackage[T1]{fontenc}

\makeatletter
\def\bstctlcite{\@ifnextchar[{\@bstctlcite}{\@bstctlcite[@auxout]}}
\def\@bstctlcite[#1]#2{\@bsphack
	\@for\@citeb:=#2\do{%
		\edef\@citeb{\expandafter\@firstofone\@citeb}%
		\if@filesw\immediate\write\csname #1\endcsname{\string\citation{\@citeb}}\fi}%
	\@esphack}
\makeatother

\begin{document}
	
	\title{Service Wrapper: a system for converting web data into web services}
	\institute{}
	\author{
	    {Naibo Wang \and
	Zhiling Luo \and
	Xiya Lyu \and
	Zitong Yang \and
	Jianwei Yin	
	}
	\email{\{wangnaibo, luozhiling, lvxy, yangzitong, zjuyjw\}@zju.edu.cn}}
	\maketitle
	
	\begin{abstract}
		Web services are widely used in many areas via callable APIs, however, data are not always available in this way. We always need to get some data from web pages whose structure is not in order. Many developers use web data extraction methods to generate wrappers to get useful contents from websites and convert them into well-structured files. These methods, however, are designed specifically for professional wrapper program developers and not friendly to users without expertise in this domain. In this work, we construct a service wrapper system to convert available data in web pages into web services. Additionally, a set of algorithms are introduced to solve problems in the whole conversion process. People can use our system to convert web data into web services with fool-style operations and invoke these services by one simple step, which greatly expands the use of web data. Our cases show the ease of use, high availability, and stability of our system.
		
	\end{abstract}

	\section{Introduction}
	
	Web services are widely used in many areas, and many program developers now tend to call the APIs offered by third-party service providers. These APIs normally receive input parameters and return formatted data such as XML or JSON documents. With the help of these web services and their APIs, the development efficiencies of programs have been greatly improved. 
	
	
	

	However, information is not always available with existing web services. Service developers usually need to spend a lot of time developing and maintaining a service, which is really a huge cost. Even so, service providers can't satisfy all the users because of the various requirements under the huge amount of data. Under many circumstances, we need to get information by extracting data from web pages. We know that the structure of HTML documents is always not in order, and web pages always contain semi-structured data or unstructured data, so a series of web data extraction methods are proposed to get structured data from HTML documents. E.g., we can use these methods to extract prices from different shopping websites to get the lowest price of a product at a moment. Other application scenarios, such as event monitoring and information retrieval, are also very useful in our daily life. Usually, we called procedures which can extract structured data from web pages \emph{wrappers}.
	
	However, it is sometimes very cumbersome to extract data from HTML documents with these wrappers. These tools usually require the help of human experts and are restricted to specific domains. For example, if someone wants to use the Lixto \cite{Baumgartner2001Visual} system to generate a wrapper of the website IMDB, he needs to know how to write the Elog rules, an internal data extraction language defined by its designer, to find and extract data from elements he wants. What's more, it is not straightforward to use the wrappers generated by Lixto. So it's of great value to directly convert the web data into callable web services without writing wrappers manually, and structured data can be retrieved from web pages by calling APIs of these web services.
	
	Since not all data in a web page are useful for us, we always need to use web block segmentation methods to get the main sections of the page. These methods can break a whole web page into many blocks and each block contains different semantic contents. We can get our desired data more easily from the related block. In general, web data extraction and web page block segmentation methods are closely related to each other.    
	
	
	In this paper, we construct a service wrapper system to convert available data in web pages into web services. We aim at generating wrappers from given web pages and transform them into web services for users in a nearly automated way. We proposed a form finding method, a block segmentation algorithm and a block sorting algorithm for the system to solve extraction problems in the conversion process. Also, we defined the data structures in the process to build our services and APIs. Users don't need to care about the internal mechanism of the generation and invocation process, they just need to finish some necessary steps, that is, specify the field attributes and select data regions they want, to get desired data then.
	
	The main contributions of this article are as follows:

		1. Construct a system that can automatically convert web data into web services without the help of human experts in the domain of web data and users can use it to generated customized web services with fool-style operations.   

		2. Define the data structures in the conversion process to help build the APIs and services.

		3. Propose and improve some algorithms in the service generation and invocation process, such as web page block segmentation and web page block sorting algorithm, to improve the automation degree of our system.
	 

	We present a demo webpage on the Internet \footnote{\url{http://service.cheosgrid.org:8076/servicewrapperdemo/index.html}} to illustrate our system, if you are interested in how we implement our system specifically, you can also get the source code of our project at Github \footnote{\url{https://github.com/ResearcherInCS/Service-Wrapper}}. 
	
	
	
	\section{Related Work}
	
	In the area of data extraction, a huge amount of wrapper generation methods are discussed in \cite{Laender2002A}. Ferrara E et al. \cite{Ferrara2014Web} have a comprehensive overview of the literature in the field of web data extraction. Many methods tend to define a type of special rules such as the Web Wrapper Factory (W4F) \cite{sahuguet1999building} which uses its declarative language to extract data from Web pages. Other tools like the Jedi \cite{huck1998jedi} and Florid \cite{may1999unified} also extract web data with their own pre-defined grammars or languages. These methods are not friendly to laypersons because they generally don't provide Graphical User Interface (GUI) to these users and thus only experts can use them.
	
	Many approaches devote themselves to simplify the data extraction process. XWRAP \cite{Liu2002XWRAP} is a supervised interactive wrapper generation tool which uses HTML DOM tree to locate elements in web pages. Other systems like the Wiccap Data Model \cite{Liu2002Wiccap} maps information from the Web into commonly perceived organization of logical concepts, and it also provides a visual tool. ANDES \cite{myllymaki2002effective} wants to solve problems occurred in the whole web data extraction process with its XML-based approach. But in general, these methods require manual postprocessors to handle errors during the interactive process and it is still hard for users without some technical background to use the generated wrappers. This is why we need to convert these wrappers into web services.
	
	Another category of methods of data extraction is based on machine learning technology. Like Phan X H et al. \cite{Xuan2005Automated} try to extract data from the web with conditional models like maximum entropy or maximum entropy Markov models. Other methods, such as Wien \cite{Kushmerick1997Wrapper} and Stalker \cite{Muslea1999A}, can automatically learn extraction rules by training examples collected from a large number of web pages. But like all the methods based on machine learning, their drawbacks are obvious such as the limited computation power and a large number of required example pages.
	
	There are also many researches on web page block segmentation. Cai D et al. \cite{cai2003vips} propose a vision-based method that simulates a human's visual perception to do the partition. Kovacevie et al. \cite{kovacevic2002recognition} divide a web page into five main sections: top, bottom, left, right and center, which can be used in web pages with relatively fixed layout.
	
	More recently, Jun Zeng et al. \cite{zeng2014web} divided a web page using the visual features of blocks, but it can only deal with blocks which are visible. Msos \cite{sarkis2015msos} used hybrid methods to segment multi screen-oriented web pages but lack of structural analysis. Then Feng H et al. \cite{feng2016web} propose a framework based on the process of analyzing and understanding web page structure. C Liao et al. \cite{liao2015event} introduce a new web page block segmentation method to extract event data from web pages whose data blocks share a similar structure. We use these methods to generate the main sections of our pages.


	A few people tried to create services with data extraction methods and wrappers, the H2W framework \cite{tatsubori2006decomposition} extracts web services from existing web applications by describing a workflow manually. Jia jing Z et al. migrate web application into web services \cite{Zhou2011Design}, but they need to collect the data from web pages and write wrappers by hand. In summary, there is still very little research on automatically converting web data into web services.
	
	\section{Problem Formulation}
	%
	
	In our work, we aim at automatically generating web services for ordinary users, and invoke these services in a very simple way. Like other data extraction tasks, we also need to handle some problems like getting data from \emph{dynamically generated Web Pages} and extracting data that are paged. In this section, we explain some terminologies and problems needed to be addressed.

	\begin{figure*}[t]
		\centering
		\subfloat[\scriptsize Amazon instance]{\label{bcasea}\includegraphics[width=.45\textwidth,height=.3\textwidth]{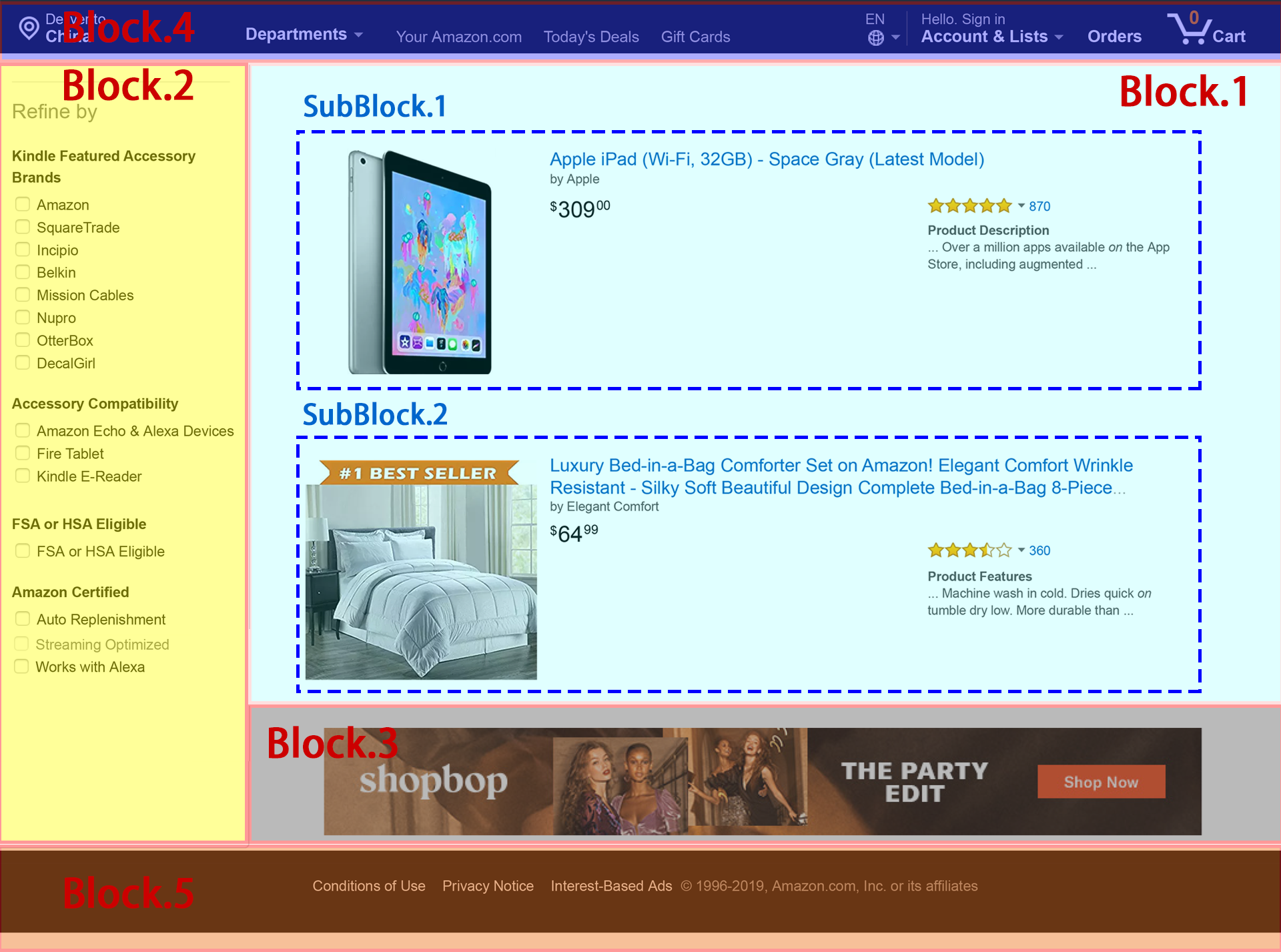}}
		\hspace{.1in}
		\subfloat[\scriptsize Taobao instance]{\label{bcaseb}\includegraphics[width=.45\textwidth,height=.3\textwidth]{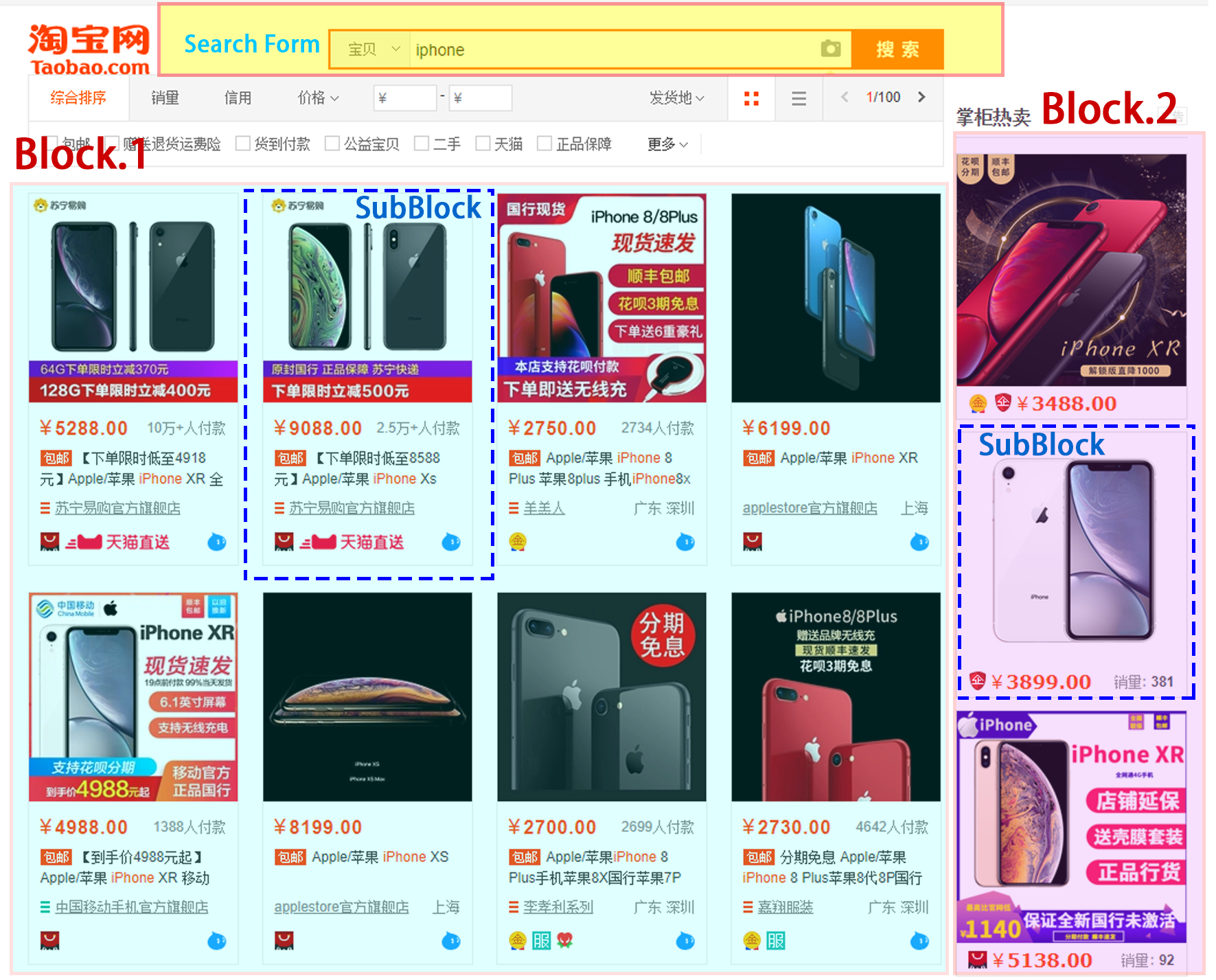}}
		\caption{Examples of block segmentation of a web page} 
		\label{blockexample} 
		\vspace{-1.5em}
	\end{figure*}

	\textbf{Block and Sub-Block:} A \text{block} is a part of a web page which contains informative data whose structure is represented as a DOM subtree of the entire page. We can divide a web page into many areas whose semantic contents are different from each other. A block usually consists of multiple \textbf{sub-blocks} whose structure is similar. As shown in Fig. \ref{bcasea}, we can divide a commodity detail page of Amazon into five blocks and Block.1 contains two sub-blocks, each sub-block contains some information of a product.
	
	\textbf{CSS Selector:} CSS selectors are patterns used to select the element(s) we want to style with CSS (Cascading Style Sheets) rules in modern browsers. We locate the elements in a web page by their CSS selectors.
	
	\textbf{Static Web Pages:} The static web pages contain information which is directly shown in the pages. We can see the data we want in a static web page when it is opened without any extra operations and extract them then. 
	
	\textbf{Dynamic Web Pages:} Dynamic web pages contain some query forms instead of direct data when they are opened. We must fill one of the query forms first with some filtering conditions and get our data in the following page. These data are related to our query conditions. For instance, as the \emph{Search Form} shown in Fig. \ref{bcaseb}, if we want to buy something online in Taobao.com, we should first enter the name of the commodity in the input box, click the search button and get the commodity list page at the end. As the commodity list page contains our desired data, it belongs to the static web page; the home page which contains the query form belongs to the dynamic web page. 
	
	Note that many web pages contain many blocks with important information. It's always up to the user to decide what blocks to be extracted. So we can just provide several candidate blocks for our user to select. 
	
	We must handle problems in the automation process as follows:
	
	\textbf{Problem 1:} How to locate and fill in the forms in a dynamic page and get valid data in the following static page.
	
	\textbf{Problem 2:} How to do the block segmentation for a variety of pages in the same way.
	
	\textbf{Problem 3:} How to decide the importance of the blocks.
	
	
	\newcommand{\Cone}{Service Extractor}
	\newcommand{\Ctwo}{Service Invoker}
	
	\section{System Framework}
	
	\begin{figure}[t] 
		\centering
		\includegraphics[width=\textwidth]{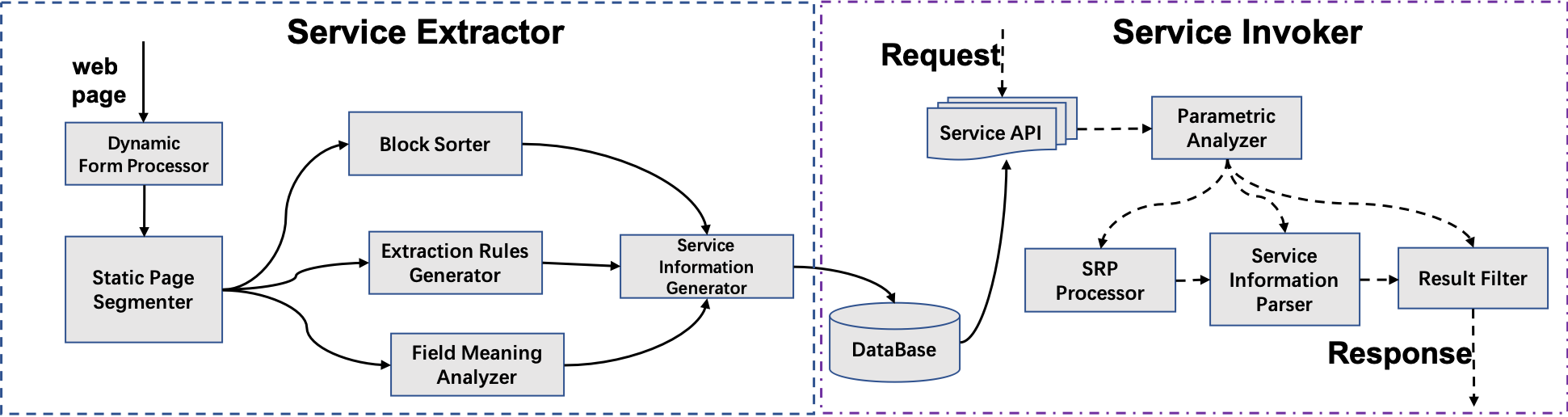} 
		\caption{Overview of Service Wrapper system} 
		\label{systemoverview} 
		\vspace{-1.5em}
	\end{figure}
	
	We introduce a system called \emph{Service Wrapper} which consists of two main components called \emph{\Cone{}} and \emph{\Ctwo{}}. The former is responsible for dynamic web pages handling, block segmentation, and service information generation. The latter is used to invoke the services (call the corresponding APIs) generated by the \Cone{}. Finally, we can get data whose structure is pre-defined by the \emph{wrapper user} with the \Cone{}. Fig. \ref{systemoverview} shows the overview of our system, the solid line and the dotted line indicate the workflows of the two components, respectively. 
	
	
	\subsection{\Cone{}}
	
	The \Cone{} is used for the service generation. It receives a web page and encapsulates its internal specified information into a web service. There are two cases we consider when extracting data from web pages: 
	
	A) Dynamic web pages. Firstly we need to find possible fillable forms on the page and provide them to our users to decide which form to fill in. After the user selects a form, he needs to enter some example values for the fields which will be automatically filled in the form by the system to get the final static web page. The following steps are the same as the static page, as shown below.
	
	B) Static web page. In this case, we can directly segment the page into blocks, sort the blocks based on their degree of importance, analyze the data structure of each block and the meaning of each field, then send these blocks to our user to decide the blocks to encapsulate finally. After slightly modifying the service info by our user, we get the web service generated by the extractor. 
	
	\newcommand{\Mone}{Dynamic Form Processor}
	\newcommand{\Mtwo}{Static Page Segmenter}
	\newcommand{\Mthree}{Block Sorter}
	\newcommand{\Mfour}{Extraction Rules Generator}
	\newcommand{\Mfive}{Field Meaning Analyzer}
	\newcommand{\Msix}{Service Information Generator}
	
	As we can see in Fig. \ref{systemoverview}, the \Cone{} contains six modules. We complete the entire work through cooperation between these modules.
	
	
	\vspace{2pt}
	\noindent \textbf{1. \Mone}
	\vspace{2pt}
	
	The \Mone{} is used to detect and extract the information of fillable forms on a web page. A form usually contains several fields and a query button. For instance, the query form in Amazon contains an input box and a search button, which are the field and the query button, respectively. In HTML5, the fields of a form are all <input> tags whose type could be \emph{text}, \emph{number}, \emph{checkbox} and so on, and the query button is also an <input> tag with the type "submit" or "button". All these elements are included in a <form> tag. When the query button is clicked, the form will be submitted in GET or POST way. However, not all the forms exist in this way, e.g., many front-end developers use the <a> tags or even a picture to represent the query button, and many forms use the AJAX technology with JavaScript code to submit the form. So it is not easy to detect the forms that do not conform to the HTML5 specification.

	
	Fortunately, we can find two common patterns of all kinds of forms: 
	
	1) We can always submit the form by clicking the query button. Since we can simulate all the click and input actions in a real web browser, we can submit the forms as long as we know where the query button is.
	
	2) Now almost all the web pages use the <input> tag to get data from their users, such as \emph{<input type='text'>}, so we can transform the input tags into query parameters in our APIs.
	
	Based on these two patterns, we can extract all the forms in a web page, record the information of these forms and their fields and query buttons.
	
	Another problem in modern web pages is that many pages are composed of many \emph{<iframe>} tags, they embed another document(s) within the current HTML document. If a form exists inside the iframe, we cannot detect it directly. So we need to switch to these iframes recursively to finish our work. In our CSS selectors, we use the mark \emph{>f>} to represent the iframe tag in a page which split the selector into two parts. 
	
	
	Alg. \ref{alfe} shows the algorithm to extract forms in a page.

		\begin{algorithm}
		\caption{Form extraction algorithm}
		\label{alfe}
		     \begin{algorithmic}[1]
			\Require The url of a web page.
			
			\State Find all the <form> tags in a web page (including the forms in <iframe> tags).
			
			\State For each form, record the information of every fillable field by its type. 
			
			\State Check the query button in order of \emph{input[type=submit]}, \emph{input[type=button]}, \emph{<button>}, \emph{<a>}, \emph{<img>} and other elements based on the click EventListeners in JavaScript.
			
			\State Find all the other elements with click EventListeners outside the <form> tags and check if the elements aside with them are fillable input boxes. If so, record them as query buttons and fields in a new form, respectively.

		\end{algorithmic}
		\end{algorithm}
	
	In step two, we record the \emph{name}, \emph{type}, \emph{value} and \emph{placeholder} attributes of the elements whose type is text, number or email; the extra checked attribute for checkbox elements; the selectIndex attribute for select and datalist elements, etc. In step three, we mean that if there is an <input> tag whose type is \emph{submit} or \emph{button}, we regard it as the query button, otherwise we choose the <button> tag as the query button and so on. We will check whether the element is bound to the click EventListeners which are widely used in JavaScript to submit forms to increase the credibility of the query buttons.

	After the form extraction, we take the screenshot of the whole page and mark all the elements with IDs generated by our system. Also, we consolidate all the records into a JSON file whose main data structures are shown in Table. \ref{J1}. The picture and file are offered to our users to decide which form and query button to choose. 
	
	\begin{table}
		\caption{Data structures of extracted forms}
		\scriptsize
		\label{J1}
		\begin{tabular}{|l|c|c|p{7.7cm}|}
		\hline
			Key name & type & parent &  Description\\
		\hline
		url & text & \emph{root} &  The url of the query page.\\
		\hline
		main\_form\_index & integer & \emph{root} & The index of selected form.\\
		\hline
		forms & array & \emph{root} & An array about extracted forms.\\
		\hline
		main\_btn\_index & integer & \emph{forms} & The index of selected button.\\
		\hline
		css\_selector & text & \emph{forms} & The position of the form.\\
		\hline
		input\_list & array & \emph{forms} & An array about the input (query) parameters.\\
		\hline
		query\_button\_list & array & \emph{forms} & An array about the query buttons\\
		\hline
		Other basic attributes & / & / & Some basic attributes about the input parameters or the query buttons such as their type, name, vaule, description, etc. The type of these attributes depends on their meaning.\\
		\hline
		\end{tabular}
	\end{table}
	
	After our user chooses a form and its query button, specifies some example values of the fields, the \Mone{} begins to automatically fill in the form with these values and click the query button. When the form is submitted, the \Mtwo{} crawls the following page.
	
	\vspace{2pt}
	\noindent \textbf{2. \Mtwo}
	\vspace{2pt}
	
	This module is to segment a given static web page into blocks with web page segmentation methods. We find that many web pages are generated based on some templates and the developer just need to fill these templates with a list of data from the back-end database to create a page. Here we segment a web page into blocks, each of which contains many sub-blocks and each sub-block shares a similar structure. We improve the web blocks segmentation algorithm in \cite{liao2015event}, as shown in Alg. \ref{blockseg}.
	
	\begin{algorithm}[t]
	    \caption{Web page block segmentation algorithm}
	    \label{blockseg}
	    \begin{algorithmic}[1]
	       \Require Root element $E_R$ of the HTML document tree. 
	       \State initalize $Q$ = \{$E$$_{R}$\}, $i$ = 0
	       \While{$Q$[$i$] is not $null$ }
	       \State $E$$_{t}$ =  $Q$[$i$]
	       \State $i$ = $i$ + 1
	       \State $C$$_{t}$ = $Childs$($E$$_{t}$)
	       \If {$E$$_{t}$ is $<iframe>$}
	       \State $switch\_to$($E$$_{t}$)
	       \State $C$$_{t}$ = $Root$($E$$_{t}$)
	       \ElsIf {$E$$_{t}$ is $<th>$}
	       \State $change\_type$($E$$_{t}$,$<td>$)
	        \EndIf
	       \If{$similar$($E$$_{t}$, $E$$_{t}$$^L$) or $similar$($E$$_{t}$, $E$$_{t}$$^{R}$)}
	       \State $mark$($E$$_{t}$)
	       \Else
	       \State $Q$ = $Q$ + $C$$_{t}$
	       \EndIf
	       \EndWhile
	    \end{algorithmic}
	\end{algorithm}
	
	$E$$_{R}$ is the root element of the HTML document; $Q$ is a squeue used for our search; $Childs$ is a function to get the child elements of a given element $E_t$; the function $Root$ is used to get the root element inside an <iframe> node; $switch\_to$ changes the context of the browser from the current page to the iframe page; $change\_type$ can change the type of a given element; the $similar$ function compares two elements based on their DOM structure and finally, the $mark$ function marks an element as a sub-block.

	At the beginning, we put $E$$_{R}$ into $Q$ and set a cursor $i$ of $Q$ starts with 0. Then we use the breadth-first search to get every element from the DOM tree of the document. As shown in line 12 to 16, in each loop, we mark current element $E_R$ as a sub-block if it is similar to its left or right brother, otherwise we will put all the child elements $C_t$ of $E_t$ into the tail of $Q$. \textbf{Finally, we merge all the sub-blocks with similar structure into one block.} Also, we handle the <iframe> tags in line 6 to 8 to get the elements inside the iframes by switching to these tags and put the root elements of them into $Q$. Another trick is to change the type of the <th> node to <td>, it's because that they represent the title and content of an HTML table respectively, so it is reasonable and convenient for us to regard their structure both as the same.

	In the end we explain how to implement the function $similar(E_1, E_2)$ to compare the similarity of two elements $E_1$ and $E_2$. We go through all the child elements of $E_1$ and $E_2$ and transform them into ordered string array $S_1$ and $S_2$. If the items of $S_1$ and $S_2$ are the same, the function returns true, otherwise it returns false.
	
	Unlike methods such as VIPS \cite{cai2003vips} which cannot find the elements that are not visible temporarily (such as some pop-up tables) and they are affected by the CSS styles and JS code, our algorithm based on the DOM tree structure can extract even invisible data and locate elements no matter where they are. Meanwhile, as we get real-time data from the browser, we can also catch the dynamically generated contents by JS code.
	
	
	\vspace{2pt}
	\noindent \textbf{3. \Mthree}
	\vspace{2pt}
	
	The \Mthree{} can sort the blocks based on their degree of importance and keep the top $n$ (pre-defined value) blocks for users to choose. After sorting, we can quickly find the block we want. Although there are various web pages, we can find some common features of blocks which are regarded as the \emph{main blocks}, like the number of data items in the main block is often higher than other blocks; the number of words and the characters in it is the highest or its size on the web page is the largest under normal conditions. Considering these factors, we propose a block sorting algorithm in Alg. \ref{blocksort}.

	\begin{algorithm}
		\caption{Block sorting algorithm}
		\label{blocksort}
		     \begin{algorithmic}[1]
			\Require Blocks, Number of blocks to keep $n$.
			
			\State Sort the blocks in descending order according to the number of lists, the number of the words and the size of the block, and put the results into three sets $BlockList$, $BlockLen$, $BlockSize$, respectively.
			
			\State Keep the first $2n$ blocks (elements) of the set $BlockList$, $BlockLen$ and $BlockSize$, then delete the remaining blocks.
			
			\State Take intersections of the three sets into set $Candidates$, then keep the top $n$ blocks in that set and get rid of the rest. 

		\end{algorithmic}
	\end{algorithm}

	Our sorting algorithm selects the block with all the three common features as a candidate main block. This makes up for the shortcomings of a single indicator, e.g., an advertisement block will not be selected for it usually contains very few words, even though its size is relatively huge in a web page. In the future, we want to import some NLP tools as auxiliaries to help us sort the blocks.
	
	\vspace{2pt}
	\noindent\textbf{4. \Mfour}
	\vspace{2pt}
	
	We need to generate the extraction rules for our \emph{\Ctwo{}} to extract data from these blocks with the \Mfour{}. Here we deal with three types of elements in a block:
	
	\textbf{1. Texts}. Text information can exist in any element in HTML documents. Notably, we should extract not only the CSS selector of the text's element but also the position of the text in the element, we call it \emph{rank}. For example, the ranks of the text "start" and "end" in the HTML code '<div>start<p>example</p>end \\ </div>' are 1 and 2 respectively for they all belong to the element <div>, and the rank of the text "example" is 1 in the element <p>.  
	
	\textbf{2. Images}. We extract the URL addresses of the images in a block. There are two types of images in HTML documents now. One is in the <img> tags; the other is the background pictures in various elements in the form of the CSS attributes. We use the type \emph{img} and \emph{background\_img} to distinguish two types of images in our rules. Also, the CSS selector of the element to which the image belongs should be recorded.

	\textbf{3. Hyperlinks}. That is, the information existing in the <a> tags, which mainly includes the address of the hyperlink and the contents inside the tag. Note that the contents also consist of the text and image information.
	
	We generate our rules for all the elements in a sub-block by their categories and put them into a JSON file, as shown below.

	\begin{quote}
	\begin{scriptsize}
	\begin{verbatim}
	{
		    "texts": [{"id":0, "rank":0, "css_selector": "html>p:nth-child(4)>h3"}],
	    "images":[{"id":1, "type":"img", "css_selector":"html>p:nth-child(4)>img"}],
			    "links":[{
               "id":2,
               "css_selector":"",  
               "texts":[{"id":3, "rank":0, "css_selector":"html>p:nth-child(4)>a"}],
               "images":[{
                           "id":4, 
                           "type":"background_img", 
                           "css_selector":"html>p:nth-child(4)>a"
                        }]
		            }]
}        
	\end{verbatim}
	\end{scriptsize}
	\end{quote}

	
	\vspace{2pt}
	\noindent\textbf{5. \Mfive}
	\vspace{2pt}
	
	The \Mfive{} is used to generate an initial name for the final output field of the API. Each output field corresponds to one of the elements in the sub-block, i.e., if a sub-block contains three elements, the corresponding API will have three output fields. We name these fields by analyzing the text belonging to the same field in all sub-blocks in this module.
	
	Most front-end developers provide some meaningful names for the elements in HTML documents by assigning values to the attributes of HTML tags, such as \emph{name, id, class, }etc. For the tables in web pages, we can directly use their header (<th> tags) to name the output fields. These are all tricks but cannot be in common use, so we use the Knowledge Graph to help us analyze the meaning of the output APIs' parameters more precisely. We send the texts to the tool and get their type by predicting the triplet \emph{(text, type,?)}. And of course, any other NLP tools may be used here as substitutes.
	
	\vspace{2pt}
	\noindent\textbf{6. \Msix}
	\vspace{2pt}
	
	This module is used to store the field information, text contents and other auxiliary data (such as \emph{service name, description, URL address, etc}) of the candidate blocks into another JSON file whose data structures are similar to the ones in Table. \ref{J1}. Likewise, we take the screenshot of the whole page and mark all the candidate blocks with their IDs. Our user will choose a block with a GUI and the corresponding field information will be displayed on the screen. The user can then modify the field names, service names and any other information in a very simple way. Finally, a customized web service is generated. 
	
	When all the work is done, the \Cone{} will create a web service based on rules generated by the above modules and store the service information into a database. Meanwhile, it generates an API address for the service, users then can call the API in the RESTful way. I.e., we call multiple APIs with the same address in different request methods (PUT, DELETE, POST, GET) to operate the web services (CRUD).
	
	\subsection{\Ctwo{}}
	
	\begin{figure}[t] 
		\centering
		\includegraphics[width=.5\textwidth]{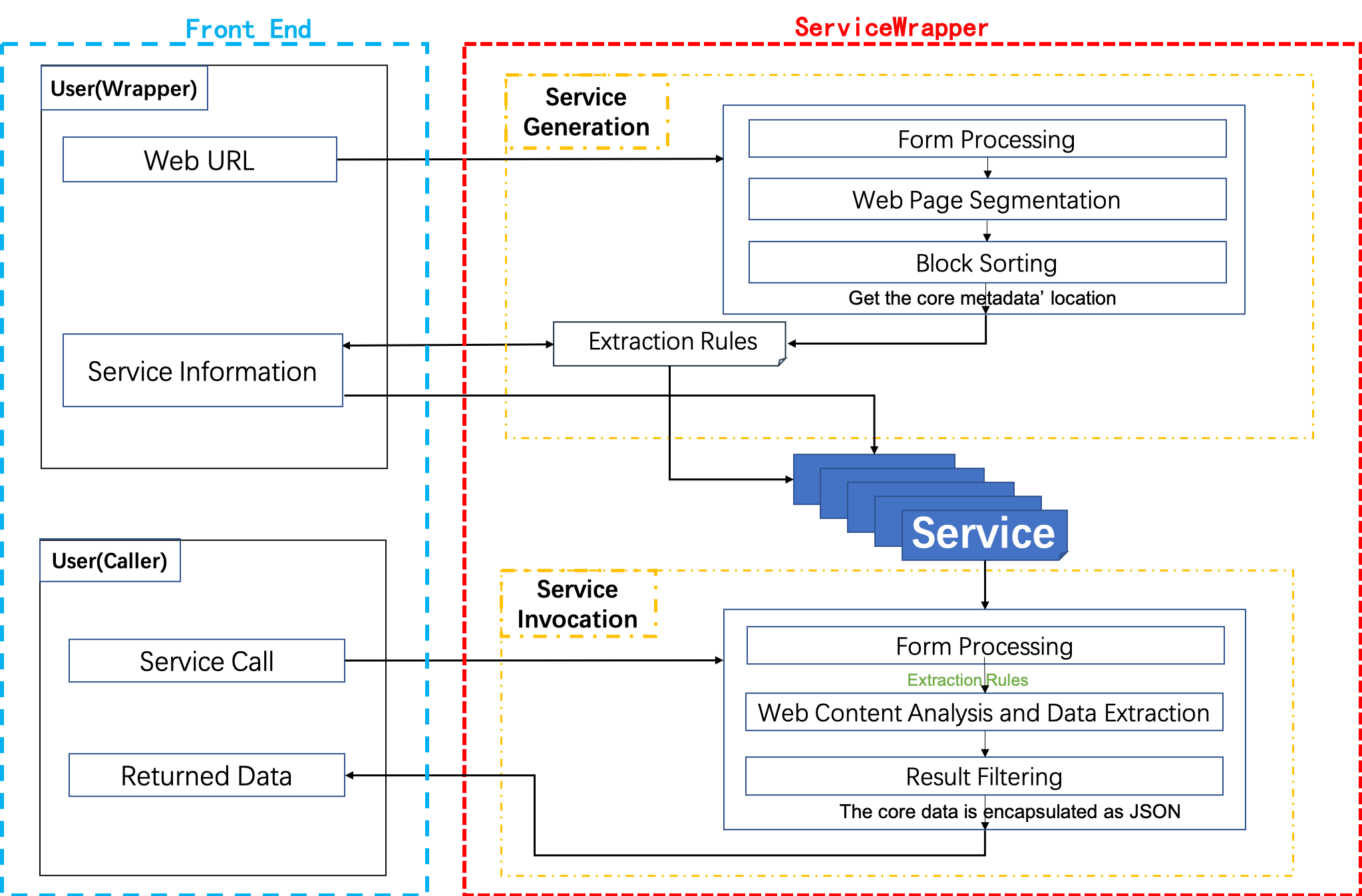} 
		\caption{Process of service generation and invocation} 
		\label{systemprocess} 
		\vspace{-1.5em}
	\end{figure}
	
	\newcommand{\Mseven}{Parametric Analyzer}
	\newcommand{\Meight}{SRP Processor}
	\newcommand{\Mnine}{Service Information Parser}
	\newcommand{\Mten}{Result Filter}
	
	The \Ctwo{} receives an API request from the caller and returns structured data designed by the wrapper user. As shown in Fig. \ref{systemoverview}, after a user sends an API request with the method GET, the \Ctwo{} accesses the corresponding service information in the database, handles the request with its modules. Analogously, the \Ctwo{} contains four modules to complete the service invocation. 
	
	\vspace{2pt}
	\noindent \textbf{1. \Mseven{}}
	\vspace{2pt}
	
	The \Mseven{} is responsible for analyzing the request parameters. We classify the request parameters into three categories:
	
	\textbf{A. System request parameters.} A service provider typically sets some common request parameters like the \emph{APIKEY} of a service. We also provide a parameter called \emph{\_\_max\_page} for all the web services in our system to extract the data that are paged.
	
	\textbf{B. Application request parameters.} They are input parameters defined by the wrapper user during service generation. These parameters correspond to the fields of a form in the dynamic web page.   
	
	\textbf{C. Filter parameters.} Sometimes the caller only wants part of the content returned by the API. Therefore he can filter out the data he doesn't want with filter parameters.
	
	The \Mseven{} will remove the parameters which do not conform to the specification and deliver the rest to the other three modules according to their categories.
	
	\vspace{2pt}
	\noindent \textbf{2. \Meight}
	\vspace{2pt}
	
	The \Meight{} is used to deal with the system request parameters. E.g., if an APIKEY is needed for the service, we can check the validity of the key here. This module serves as a slot for the service provider and could be redeveloped.
	
	
	\vspace{2pt}
	\noindent \textbf{3. \Mnine}
	\vspace{2pt}
	
	We handle the application request parameters here. Firstly, the \Mnine{} maps the keys of input parameters to the actual fields of a specified form based on the rules generated by the \Mone{}, and fill in and submit the form with those keys and values. Then it locates specified elements and gets their data on the following static web page with the extraction rules. Finally, this module transforms those data into structured data with a pre-defined format from the \Msix{}. We won't repeat the details here for they are the same as before.
	
	\vspace{2pt}
	\noindent \textbf{4. \Mten}
	\vspace{2pt}
	
	The \Mten{} helps the caller to filter out some items from the output of the last module with some filtering conditions. For instance, if there is an output parameter called \emph{price} in a service, we can use the key-value pair "price=35" to get the products whose price is \$35 from the whole set. And also, they can use arithmetic operators defined by our system. What's more, the parameters may be in nested structures, e.g., the key "pc.price" refers to element \emph{price} nested in the link element \emph{pc}. After the filtering is complete, the module will return the final result, that is, the output of the API.
	
	Meanwhile, as data are always paged because of their large quantity, we need to come up with a method to extract them from multiple pages at once. We use an algorithm to search for elements containing text like 1, \emph{next page}, and \emph{last page} in the depth-first search way and get their CSS selectors. As the position of these elements usually does not change, we can keep extracting the data by clicking the \emph{next page} tags until we are on the last page. Remember that the parameter \emph{\_\_max\_page} specifies the maximum number of pages to be extracted.

	The whole process of service generation and invocation is shown in Fig.\ref{systemprocess}. Our system can be seen as a black box, and what users do in the front end is very simple.
	
	
	\section{Case Study}
	
	\begin{figure*}[t]
		\centering
		
		\subfloat[\scriptsize Form Detection and Block Segmentation]{\label{casea}\includegraphics[width=.35\textwidth,height=.3\textwidth]{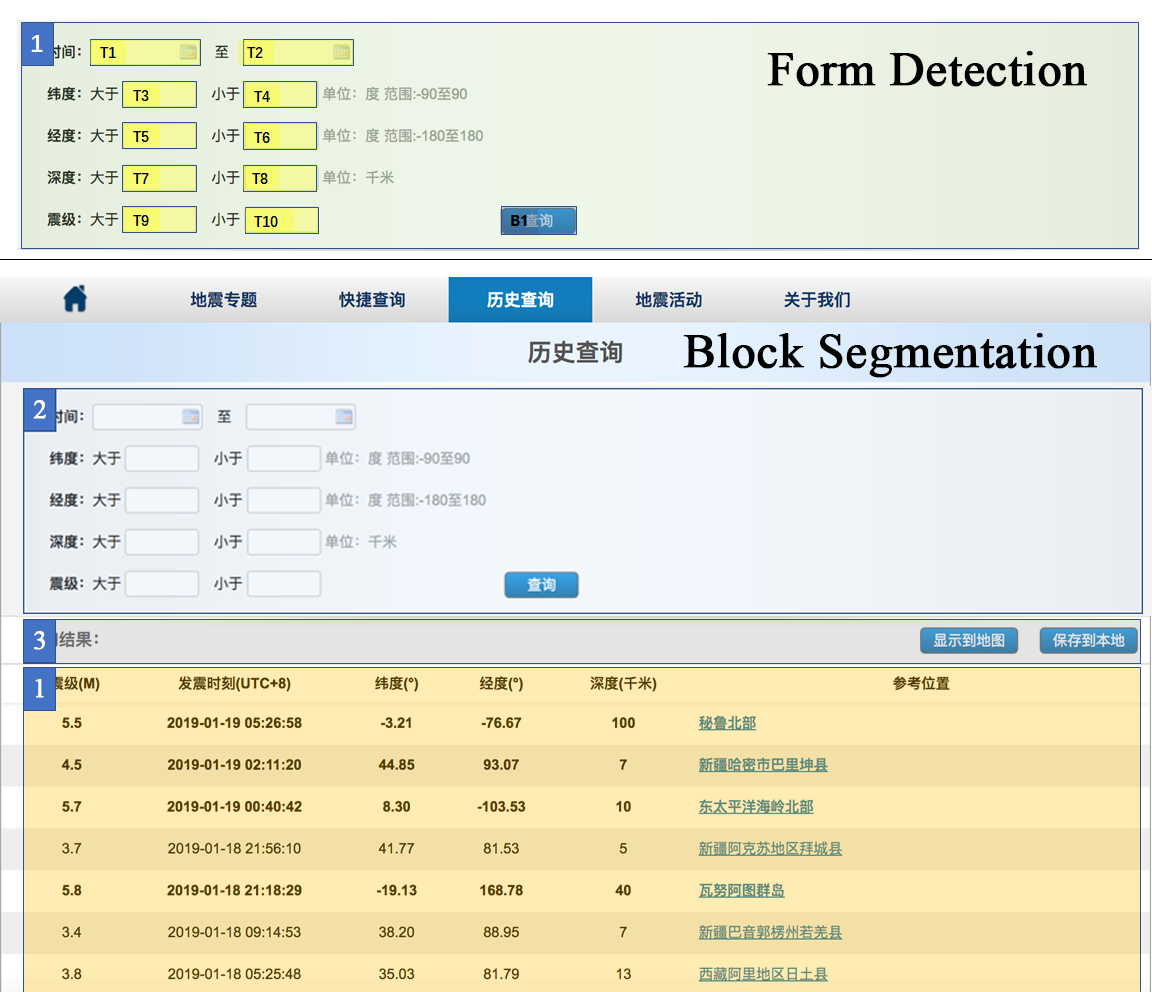}} 
		\subfloat[\scriptsize Parameters Configuration and Block Selection]{\label{caseb}\includegraphics[width=.3\textwidth,height=.3\textwidth]{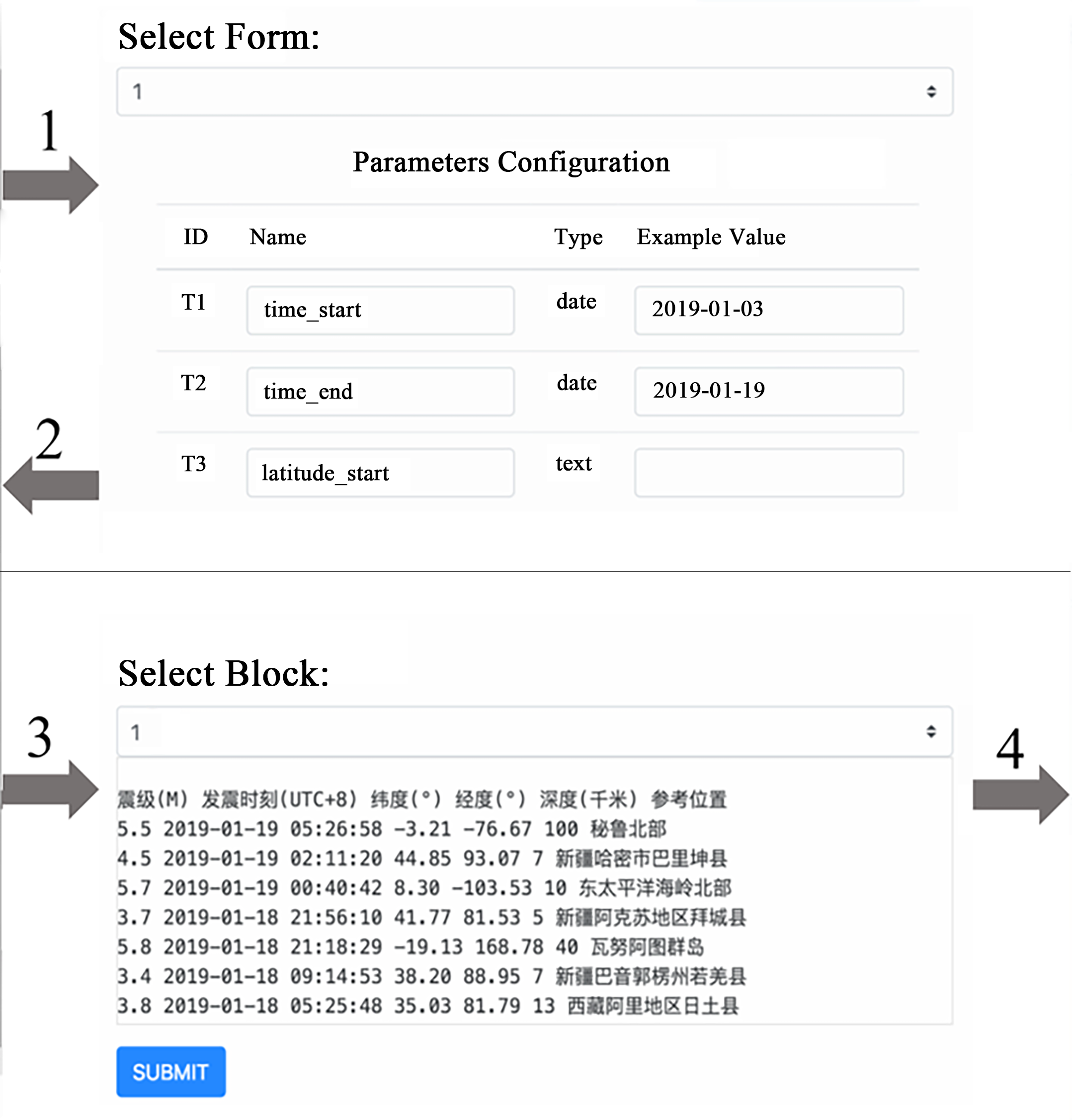}}
		\subfloat[\scriptsize Generated Service Information]{\label{casec}\includegraphics[width=.35\textwidth,height=.3\textwidth]{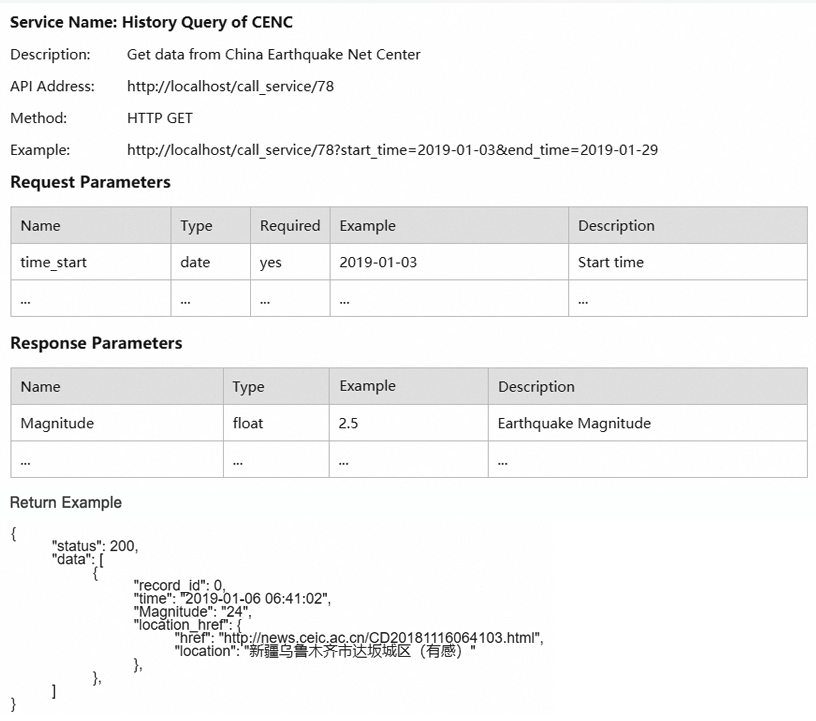}}
		\caption{An instance of the service generation by the Service Wrapper}
		\label{caseexample}
		\vspace{-1.5em}
	\end{figure*}

	\subsection{History query page in the China Earthquake Network}

	We use the history query page in the China Earthquake Network to illustrate the instance of the dynamic page with form and the following static page. The page provides a form including some query fields such as latitude, longitude, etc. After we enter our query conditions in the fields and click the query button, the seismic data we want will be displayed in the table below the form. Our goal is to convert this page into a web service with our system.
	
	\vspace{2pt}
	\noindent \textbf{1. Service Generation}
	\vspace{2pt}
	
	Fig. \ref{caseexample} shows the whole conversion process whose arrows represent the order of execution. At first, we input the URL (\url{http://www.ceic.ac.cn/history}) into the system, and our system will convert the page into a web service as follows:
	
	1) Form Detection. As shown in the top of Fig. \ref{casea}, the \Mone{} detects the query form in the page and marks it in area 1. All the input fields are marked from number \emph{T1} to \emph{T10}, and the only candidate query button is marked as \emph{B1}. As shown in the top of Fig. \ref{caseb}, we then can select the form and corresponding query button, configure the parameters and enter some example values based on this figure. Here we query the seismic data from date \emph{2019-01-03} to \emph{2019-01-19}. Our system will automatically fill in the form with these example values and get the following static web page. 
	
	2) Block Segmentation. The \Mtwo{} divides the static page into three blocks, they are all numbered. Without any manual intervention, we just need to wait for the segmentation process to complete. As shown in the bottom of Fig. \ref{casea}, the first block whose serial number is 1 contains the seismic data we want in the form of a table. 
	
	3) Block Selection and Parameters Configuration. As shown in the bottom of Fig. \ref{caseb}, we select the list block (block 1) based on the image and the main text information it contains. We then configure the output parameters for the block in the same way, as shown in the top of Fig. \ref{caseb}.
	
	4) Service Generation. Finally, we configure some basic information about the service, such as name, description, etc. After we click the \emph{submit} button, the \Cone{} generates a web service successfully, as shown in Fig. \ref{casec}.

	\vspace{2pt}
	\noindent \textbf{2. Service Invocation}
	\vspace{2pt}
	
	We can see that the \Cone{} provides an API address (http://localh\\ost/call\_service/78) for the generated service in Fig. \ref{casec}. We can get desired data by calling the API in a RESTful way with the request parameters. Remember that we have three types of request parameters. For instance, if an URL is:
	
	\vspace{0.5em}
	\small
	http://localhost/call\_service/78?start\_time=2019-01-01\&end\_time=2019-01-18\\\&Magnitude=20\&key=a1f6de13
	\normalsize
	\vspace{0.5em}
	
	We mean that we want to get the seismic data from  \emph{2019-01-01} to \emph{2019-01-18} and get data whose magnitude is \emph{20} because that the \Mten{} will filter out irrelevant data according to parameter \emph{Magnitude}. Finally, we specified a key \emph{a1f6de13}, it is a system request parameter to get the service invocation permission. During the invocation process, we only need to specify the input parameters to get the desired structured data. All the work of data extraction and transformation is done automatically by the \Ctwo{}. Our final result is shown at the bottom of Fig. \ref{casec} is in the format of JSON.
	
	
	\subsection{Douban movie list page}

	\begin{figure}[t] 
		\centering
		\includegraphics[width=.4\textwidth]{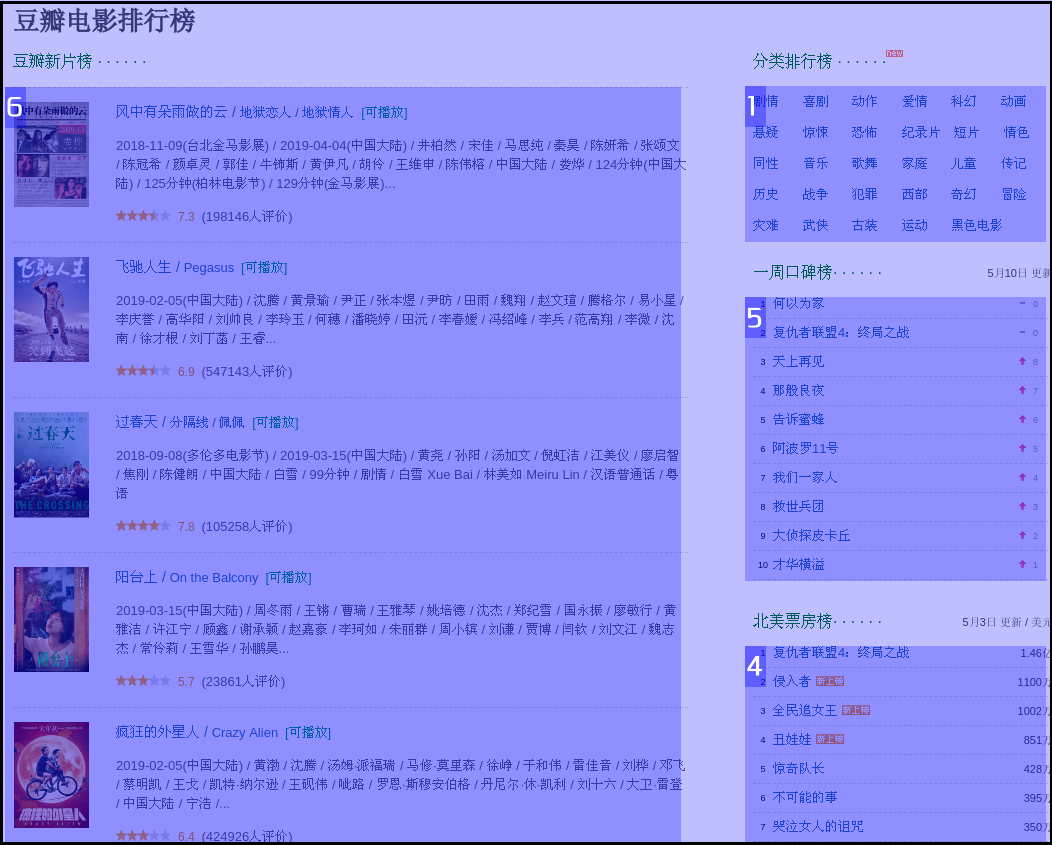} 
		\caption{Instance of block segmentation of Douban movie list page} 
		\label{douban} 
		\vspace{-1.5em}
	\end{figure}

	Douban.com, like IMDB, is a Chinese online database of information related to films, books, movies, etc. We choose the movie list page in Douban.com as another case to show how to extract the movie information from multiple blocks. Fig. \ref{douban} is a part of the screenshot after the block segmentation of the page which contains four blocks. Block 6 contains the most popular new movies in China whose information includes pictures, names, rating scores, descriptions and so forth. Block 4 is the one week of word of mouth movie list and Block 5 is the North American box office list. Here we want to get these three parts of data in the form of services. 
	

	\vspace{2pt}
	\noindent \textbf{1. Service Generation}
	\vspace{2pt}

	We need to select all three blocks at the same time in the service generation phase here. Similar to the last case, we enter the URL (\url{https://movie.douban.com/chart}) of the page, skip the form filling process by clicking the \textit{skip form} button and select block 4, 5, 6 by clicking the \emph{add section} button as in the bottom of Fig. \ref{caseb}. Then we configure the blocks' parameters at the same time. The whole generation process is almost the same as the last case. After clicking the submit button, the service is successfully generated by our system.

	\vspace{2pt}
	\noindent \textbf{2. Service Invocation}
	\vspace{2pt}

	Likewise, we invoke the service the same way, i.e., we get the movie data by calling the API of the service with method GET. Here we can get the real-time popular movies as well as their image link addresses, descriptions and other useful info. Meanwhile, we can get info from other lists such as the North American box office list. We can also sift these movies through their attributes. All the services are invoked in the same simple way with our system.

	As we can see, no code or configuration file writing, no technical means is required throughout the whole generation and invocation process. Compared with other tools, we just need to configure something subjective such as what form or blocks to select, what's the names of parameters, etc. Without our system, the process has to be completed in whole or in part by professionals. Our system surpasses other methods in terms of convenience and automation. We believe our users will do more interesting things with our system.

	\section{Conclusion}
	
	In this paper, we construct a system called Service Wrapper to convert available data in web pages into web services. Our implementation is based on web data extraction and block segmentation technologies. After the generation of a web service, our users can call the APIs of the service just like other web services. In addition to the parts that must be manually processed, our system is basically fully automated. Our system can handle both static web pages and dynamic web pages. The high availability and stability of our system are shown in our cases.

	
	
	\bibliographystyle{splncs04}
	\bibliography{IEEEexample}
	
\end{document}